\documentclass[12pt]{article}
\usepackage{a4wide}
\usepackage{latexsym}
\usepackage{epsfig}
\begin{document}

\thispagestyle{empty}

\begin{flushright}
NIKHEF/2003-003\\
BI-TP 2003/08
\end{flushright}
\vspace{1.5cm}
\begin{center}
  {\Large \bf The pion form factor in improved lattice QCD}\\[.3cm]
  \vspace{1.7cm}
  {\sc J. van der Heide and M. Lutterot}\\
  {\it National Institute for Nuclear Physics and High-Energy Physics
    (NIKHEF)\\ 1009 DB Amsterdam, The Netherlands}\\
  \vspace{1cm}
  {\sc J.H.\ Koch, }\\
  {\it NIKHEF and Institute for Theoretical Physics,
    University of Amsterdam, Valckenierstraat 65, 1018 XE Amsterdam, 
    The Netherlands}\\
  \vspace{1cm}
  {\sc E.\ Laermann}\\
  {\it Fakult\"at f\"ur Physik, Universit\"at Bielefeld,
    D-33615 Bielefeld, Germany}\\
\end{center}
\vspace{2cm}

\begin{abstract}
We calculate the electromagnetic form factor of the pion in lattice 
gauge theory. The non-perturbatively improved Sheikoleslami-Wohlert 
lattice action is used together with the ${\cal O}(a)$ improved 
current. The form factor is compared to results for other choices for 
the current and features of the structure of the pion deduced from 
the 'Bethe-Salpeter wave function' are discussed.\\
\\
PACS Indices: 11.15.Ha, 12.38.Gc
\end{abstract}
\vspace*{\fill}

\section{Introduction}
The pion as the simplest particle with only two valence quarks has 
been the subject of many studies. Global features of the pions - 
their charge and spin -  are easily incorporated in model calculations. 
The form factor, which directly reflects the internal structure of 
this elementary particle, is clearly an important challenge. Many 
earlier calculations are based on {\it ad hoc} models that model QCD or 
sum over selected subsets of Feynman diagrams. However, the most 
reliable approach, in particular when addressing non-perturbative 
features as the electromagnetic form factor at intermediate momentum 
transfers, is the use of lattice QCD. The first lattice results 
were obtained by Martinelli and Sachrajda \cite{Martinelli}, which was 
followed by a more detailed study by Draper {\it et al.} \cite{Draper}, 
who showed that the form factor obtained through lattice QCD with the 
Wilson action could be described by a simple monopole form as suggested 
by vector meson dominance \cite{Williams}. Below, we extend these early 
studies in two ways. We use an improved lattice action and an 
${\mathcal{O}}(a)$ improved electromagnetic current operator. 
Furthermore, we also extend the calculations to lower pion masses than 
achieved before. Several features of the internal structure of the pion 
have been obtained previously \cite{Chu,Hecht,Gupta,Lacock,Schmidt} by 
calculating the 'Bethe-Salpeter wave function', which can be used to 
estimate the relative separation of the quark-antiquark pair in the 
pion. We also use this approach and compare its predictions to the 
results of our direct calculation of the pion form factor.

\section{The method}
In comparison to earlier lattice calculations of pion properties, the 
major difference of our approach is the systematic reduction of the 
discretisation error in the calculation of the matrix elements. We use 
the non-perturbatively $\mathcal{O}(a)$ improved \cite{Luscher} clover 
action \cite{Sheikholeslami} and the corresponding $\mathcal{O}(a)$ 
improved current \cite{Sachrajda,Luscher2,Sommer}.

Using this action, we proceed analogous to \cite{Draper} and 
calculate the two- and three-point correlation functions for the pion.
Projecting onto definite pion three-momentum, the two-point function is
\begin{equation}
G_2 (t, {\bf p}) = \sum_{\bf x} \left<\phi({t, \bf x})\; \phi ^{\dagger}
(0,{\bf 0})\right> \; e^{i\; {\bf p}\cdot {\bf x}} \; ,
\end{equation}
where $\phi$ is the operator projecting on a state with the pion 
quantum numbers. Below, we will consider a $\pi^+$ meson,
consisting of a $u$ and $\bar d$ quark. Neglecting all spin, colour and 
flavour indices, this operator is
given by \begin{equation}
\phi (x) = {\bar \psi} (x)\; \gamma_5\; \psi (x) \; .
\end{equation}

In the three point function, which yields the desired form factor, we 
project onto specific initial and final pion three momenta, ${\bf p}_i$ 
and ${\bf p}_f$ 
\begin{equation}
G_3 (t_f, t; {\bf p}_f , {\bf  p}_i) = \sum_{{\bf x}_f, {\bf x}} \;
\left<\phi (x_f) \; j_4 (x) \; \phi^{\dagger} (0)\right>  e^{-i\; {\bf
p}_f \cdot ({\bf x}_f - {\bf x}) \; - i \; {\bf p}_i \cdot {\bf x}} \;. 
\end{equation}
This function involves the electromagnetic current operator $j_\mu$; 
since here we use only the component $\mu = 4$, we do not include a 
$\mu$-index in the definition of the three point function.

It is well known that the local current,
\begin{equation}
j^{L}_\mu (x) = {\bar \psi (x) }\; \gamma_\mu \; \psi (x),
\end{equation}
is not conserved on the lattice. The conserved Noether current that 
belongs to our action,
\begin{equation}
j^{C}_\mu=\kappa\left({\bar\psi(x)}(1-\gamma_\mu)U_\mu(x)\psi(x+{\hat
\mu})-{\bar\psi(x+{\hat\mu})}(1+\gamma_\mu)U^{\dagger}_\mu (x)\psi (x) 
\right),
\end{equation}
is identical to the conserved current for the Wilson action and still
contains corrections of  ${\mathcal{O}}(a)$ at $Q^2 \neq 0$; we use
$Q^2=-q^2$, where $q$ is the four momentum transfer to the pion.

The conserved and improved vector current $j^I_{\mu}$  is of the form 
\cite{Sachrajda,Luscher2,Sommer} 
\begin{equation}
j^I_{\mu} = Z_{V} \, \{ j^{L}_\mu (x)  + a \, c_V \, \partial _\nu \, 
T_{\mu \nu} \} \;, 
\end{equation} 
with 
\begin{eqnarray} 
T_{\mu \nu} &=&  {\bar \psi}(x)\; i \; \sigma_{\mu \nu}  \; \psi (x) 
\;,\\ \nonumber
Z_V &=& Z_V^0 \: (1+ a\,b_V\,m_q) \; .
\end{eqnarray}
The bare-quark mass is obtained from:
\begin{eqnarray}
a\;m_q=\frac{1}{2}\left(\frac{1}{\kappa}-\frac{1}{\kappa_{c}}\right)
\end{eqnarray}
where $\kappa_{c}$ is the kappa value in the chiral limit and $a$ the 
lattice spacing. For our simulation we use $\kappa_{c}=0.13525$ 
\cite{Bowler}.
The constants in $j^I_\mu$ are determined such that the matrix element 
of the current operator receives no correction to ${\mathcal{O}}(a)$. 

\section{Details of the calculation}
Our calculations were carried out in the quenched approximation on an
${N_\sigma}^3 \times N_\tau = 24^3 \times 32$ lattice and based on a set 
of $100$ configurations for the link variables at $\beta = 6$ and 
$c_{SW}=1.769$ \cite{Luscher}. After an initial thermalisation of 2500 
sweeps, we obtained configurations at intervals of 500 sweeps. Each 
sweep consists of a pseudo-heatbath step with FHKP updating in the 
$SU(2)$ subgroups, followed by four over-relaxation steps. In contrast 
to the Dirichlet conditions in \cite{Draper}, we impose anti-periodic 
boundary conditions on the quarks and periodic boundary conditions on 
the gluons. Three values of the hopping parameter $\kappa$ were used
\begin{equation}\kappa_1= 0.1323 \; , \kappa_2 = 0.1338 \; , {\rm and}
\;\; \kappa_3 = 0.1343.
\end{equation}
For the improved current, we use the parameters $Z_V^0$, $b_V$ and 
$c_V$ as determined by Bhattacharya {\it et al.} \cite{Bhattacharya}.

Conservation of the total charge generated at the source at $t=0$ 
provides a test \cite{Draper} for our calculation, relating the 
$\mu = 4$ component of the three-point function for ${\bf q}=0$ to the 
two-point function. For our periodic boundary conditions, it reads
\begin{equation}
G_3(t_f, t; {\bf p}, {\bf p}) - G_3(t_f, t'; {\bf p}, {\bf p}) = G_2 
(t_f,{\bf p}) \; ,
\label{eq:chargecons}
\end{equation}
where $t_f < t' < N_\tau$. We find that all configurations we use 
each satisfy this condition to at least 1 ppm.

For the results discussed below, we chose the pion three-momenta such 
that $|{\bf p}_i|^2= |{\bf p}_f|^2= 2$ in units of the minimal momentum 
$\frac{2\;\pi}{a\;N_\sigma}$ for our lattice. This guarantees for the 
elastic pion form factor that $E_f - E_i = q_0 = 0$; it greatly 
simplifies the kinematic factors appearing in the three-point function. 
Different values for the three momentum transfer ${\bf q}$ were obtained 
by varying the relative orientation of the initial and final pion 
momenta. 

In order to improve the projection onto the ground state, we smeared the 
pion operator at the sink in $G_2$ and $G_3$ by the method proposed in 
\cite{Lacock}. We found that a quark-antiquark distance $R = 3$ 
works best. The quark-antiquark pair was connected by APE smeared gluon 
links at smearing level $4$ and relative weight $2$ between straight 
links and staples.

To extract the desired information from our numerical results, we assume 
the two-point function of the pion to have the form
\begin{equation}
G_2 (t,{\bf p}) = \sum_{n=0}^1{\sqrt {Z_R^n({\bf p})\: Z_0^n({\bf p})}} 
\; e^{-\; E^n_{\bf p}\;\frac{N_\tau}{2} }\;  \cosh \{E^n _{\bf p}\: (
\frac{N_\tau}{2} - t)\} \; ,
\end{equation}
including the contribution of the ground state (n=0) and a first excited 
one (n=1). The $Z^n_R$ denote the matrix elements,
\begin{equation}
Z^n_R({\bf p}) \equiv |\left<\Omega | \phi_R | n, {\bf p} \right>|^2 \;,
\end{equation}
and $E^0_{\bf p}$, $E^1_{\bf p}$ are the energies of ground and excited 
state, respectively; the subscript $R$ indicates the operator smearing.

The three-point function is parametrised as 
\begin{eqnarray}
G_3(t_f,t;{\bf p}_f,{\bf  p}_i) = F(Q^2) \; \sqrt {Z_R^0 ({\bf p}_f)\; 
Z_0^0 ({\bf p}_i) } e^{- E^0_{{\bf p}_f }\; (t_f - t)\,  - E^0_{{\bf 
p}_i}\; t } \nonumber\\
+\left\{\sqrt {Z_R^1 ({\bf p}_f) Z_0^0 ({\bf p}_i)}\left<1,{\bf p}_f| 
j_\mu (0) |0,{\bf p}_f \right>e^{-\, E^1_{{\bf p}_f} \, (t_f - t)  -\, 
E^0_{{\bf p}_i} \, t} \;+ (1 \leftrightarrow 0) \right\}\;.
\end{eqnarray}
Effects involving, for example, the production of pion pairs, as well as
 'wrap around effects' due to the propagation of states beyond $t_f$ are 
exponentially suppressed ($ < \mathcal{O}$$(e^{-5})$); similarly, an 
elastic contribution from the excited state was estimated to be of the 
order of $1\%$ or less. All these effects are not reflected in our 
chosen parametrisation.

All parameters in the 2- and 3-point functions - energies $E$, 
$Z$-factors and the form factor $F(Q^2)$ - were fit simultaneously to 
the data from all configurations. For the three-point function, we chose 
$t_f = 11$ and let the current insertion time $t$ vary from $0$ to $10$. 
For maximum spatial symmetry, all values corresponding to the same value 
$|{\bf p}|$ in the two-point function and all ${\bf p}_{i,f}$ yielding 
the same ${\bf q}$ in the three-point function were combined for the fit. 
The value for the parameters and their error in these simultaneous fits 
was obtained through a single elimination jackknife procedure. Since we
satisfy eq.~(\ref{eq:chargecons}) to high accuracy, we show F(0)=1 in the
results below instead of using the result from a fit at $Q^2=0$, which 
would be less accurate in this case.

\section{Results}
\begin{figure}
\begin{center}
\includegraphics[height=120mm,angle=270]{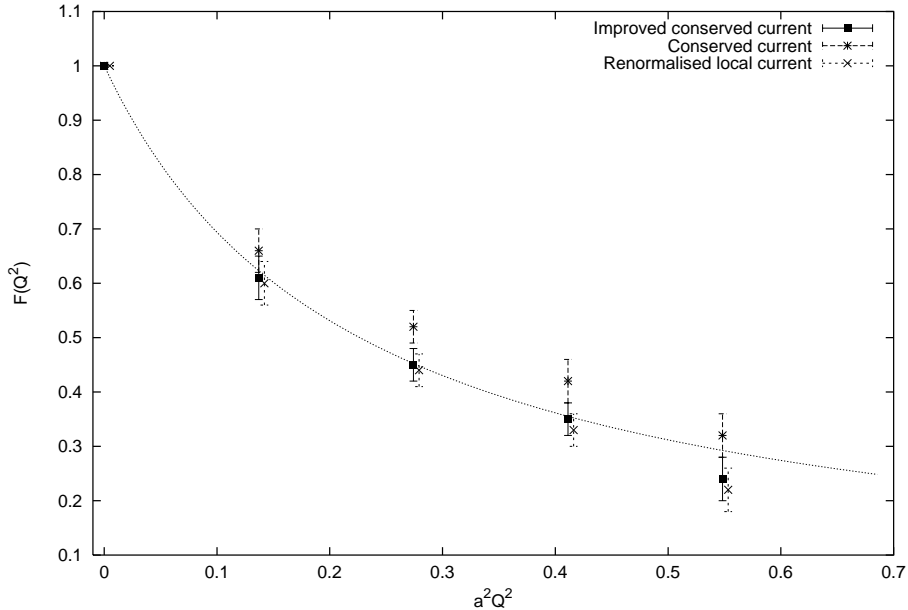}
\end{center}
\caption{{\em Form factors for the three currents with lowest quark mass. 
Solid line: monopole form with $m_{\rho}$ taken from literature (see 
text). Data for the local current shifted horizontally for 
clarity.}}
\label{fig:Icl_comp}
\end{figure}

Our method to extract the pion form factor is non-perturbatively improved 
in two respects: we use an improved action and an improved current 
operator. We can get an impression of the importance of the latter effect 
by comparing the conserved Noether current corresponding to the improved 
action with the improved current (which is also conserved). The results 
\footnote{In all our results we only show the statistical errors.} are 
shown in Fig.\ref{fig:Icl_comp} for the lightest of our three 
quark masses. The form factor from the improved current is systematically 
lower than the one from the conserved current. The difference grows with 
$Q^2$ and reaches about $25 \%$ at the largest momentum transfer 
considered here.

The structure of the improved current can be further understood by 
comparing the improved current  to the renormalised local current,  
\begin{equation}
j_\mu^{L, R} \equiv Z_V\;\bar{\psi}\;\gamma_{\mu}\;\psi\; ,
\end{equation}
which is not conserved. For our kinematics with $q_0 = 0$, we can also
extract a form factor from $j_4^{L}$. It is also shown in 
Fig.\ref{fig:Icl_comp} and can be seen to lie very close to the 
improved current. This means that the contribution of the term 
proportional to $c_V$ in $j^{I}_4$ is very small. Closer inspection
shows that while the matrix element of the tensor term can become 
almost comparable to that of the $\gamma_4$ term, the overall tensor 
contribution is reduced by the small coefficient $c_V$. Similar 
statements also hold for our two additional $\kappa$-values.
\begin{figure}
\begin{center}
\includegraphics[height=120mm,angle=270]{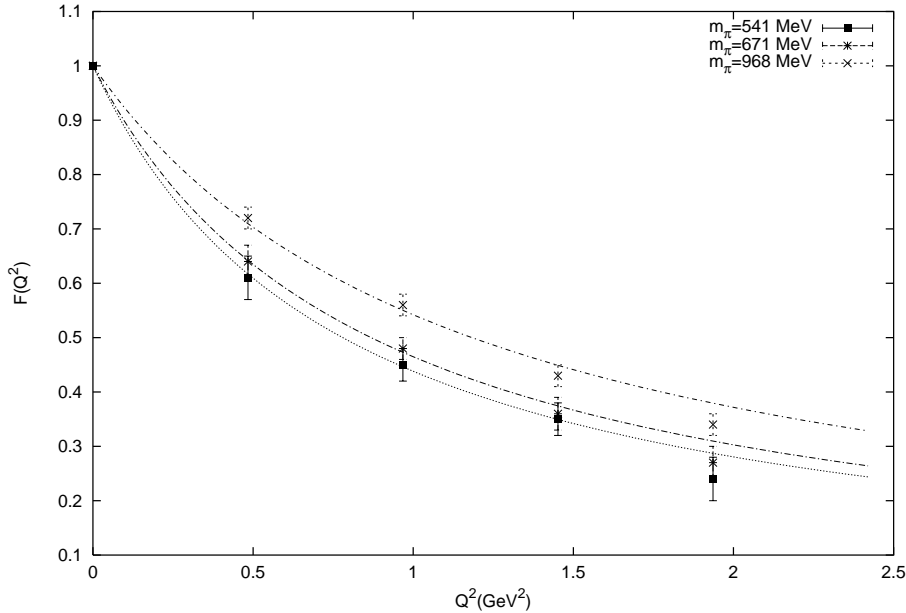}
\end{center}
\caption{{\em Form factors as a function of $Q^2$ for the three pion 
masses. Curves: monopole fits to the data.}}
\label{fig:ff_VMD}
\end{figure}

It is worth mentioning that with the $Z_V$, $c_V$ and $b_V$ values taken 
from \cite{Bhattacharya}, and performing a fit at $Q^2=0$ we obtained 
$F^I/F^C=1$ to better than $1\%$ with a statistical error 
of about $5 \%$.

In the previous study \cite{Draper} of the pion form factor, where the 
Wilson action was used, the results were compared to a monopole form 
factor
\begin{equation}
F(Q^2) = \{ 1 + \frac{Q^2}{m_\rho^2}\}^{-1} \;,
\end{equation}
a form suggested by vector meson dominance. We also show in 
Fig.\ref{fig:Icl_comp} a monopole form factor using the value for the 
$\rho$-mass obtained by interpolating the results from \cite{Bowler} 
which uses the same action as we do. This monopole form factor describes 
our results for the improved current at all but the highest $Q^2$ very 
well. As in \cite{Draper}, we observe that the conserved current lies 
consistently above the monopole form factor. A similar behaviour was 
found also for our other two $\kappa$-values.

In Fig.\ref{fig:ff_VMD} we show our results for improved form factors 
for all three values for $\kappa$. The corresponding quark- and pion 
masses are given in Table \ref{table1}. The form factors systematically 
decrease with decreasing pion mass. The form factors for the two 
lightest pions, with\footnote{For definiteness, we have taken the 
lattice spacing $a = 0.105\; {\rm fm}$ from \cite{Edwards}.} $m_\pi$=541 
and 671 MeV, come very close together. As can be seen, the statistical 
error of the extracted form factors grows as the 
quark mass decreases. Nevertheless, we are still able to obtain 
conclusive results for the highest $\kappa$. The corresponding pion mass 
of 541 MeV is substantially lower than in the previous work, where 
$m_\pi \sim$ 1 GeV. Given that we have only three data sets, we have not 
attempted to extrapolate our improved form factor to the physical 
pion mass.

We also fitted our results for the improved form factors to a monopole 
form factor. In doing so, we omitted the highest momentum data point and 
extracted in each case a vector meson mass, $m_{V}$, shown in 
Table \ref{table1}. They are close to the values for $m_\rho$ 
taken from interpolations to literature data \cite{Bowler}.

In examining the two-point Green function for various quark-antiquark 
distances, we also obtain the 'Bethe-Salpeter wavefunction',
\begin{eqnarray}
\Phi_{BS}(R) = \sqrt{\frac{Z_R^0(\bf 0 )}{Z_0^0(\bf 0 )}}.
\end{eqnarray}
Following the procedure in \cite{Hecht,Schmidt},
we obtain $\left<r^2\right>^{\frac{1}{2}}_{BS}$, 
shown in Table \ref{table1}.
\begin{table}
\begin{center}
\begin{tabular}{|l|c|c|c|c|c|c|}
\hline
$\kappa$ & $m_q$ & $m_{\pi}$ & $m_{\rho}$ 
& $m_{V}$ & $\left<r^2\right>^{1/2}_{BS}$ & 
$\left<r^2\right>^{1/2}_{FF}$\\
\hline
$0.13230$ & $0.082$ & $0.515(2)$ & $0.625(5)$ & 
$0.597(14)$ & $2.530(2)$ & $4.23(10)$\\ 
$0.13380$ & $0.040$ & $0.357(2)$ & $0.513(5)$ & 
$0.496(15)$ & $2.615(2)$ & $4.94(15)$\\
$0.13430$ & $0.026$ & $0.288(2)$ & $0.476(7)$ & 
$0.470(19)$ & $2.629(2)$ & $5.21(21)$\\
\hline
\end{tabular}
\end{center}
\caption{{\em Masses and RMS-values for the different kappa values, in 
lattice units. The $\rho$-mass has been taken from \cite{Bowler}.}}
\label{table1}
\end{table}
These RMS-radii are compared to the values extracted from the low-$Q^2$ 
behaviour of the form factor,
\begin{equation}
\left. \frac{dF(Q^2)}{dQ^2}\right|_{Q^2=0}=-\frac{1}{6}\left<
r^2\right>_{FF} = -\frac{1}{m_{V}^2}
\end{equation}
where in the last step we have assumed a monopole form and use the fitted
parameter $m_V$. In agreement with the findings of \cite{Chu,Hecht,Gupta}, 
we see that the Bethe-Salpeter predictions are very insensitive to the 
value of the quark or pion mass. However, it is well known \cite{Gupta} 
that the information that can be obtained from the Bethe-Salpeter approach 
as described above is only an approximation. It assumes, in the extraction 
of $\left<r^2\right>$, that the center of mass of the pion is always 
halfway between the valence quark and antiquark, not allowing for the 
motion of the gluons. The extraction of $\left<r^2\right>$ from the 
calculated pion form factor does not involve this restriction for the 
valence (anti-)quark motion. As can be seen, the more reliable 
determination from $F(Q^2)$ leads, as expected, to a larger radius. 
Moreover, this radius shows a substantial dependence on the mass.

We have presented here the first calculation of the electromagnetic 
form factor of the pion based on an $\mathcal{O}(a)$ improved action and 
the concomitant improved vector current. This is seen to lead to 
significant changes in the prediction for the internal structure of the 
pion. We observe a decrease of the form factor for decreasing pion mass, 
which in turn leads to an increase of the RMS-radius. This 
mass-dependence of the radius is not seen in the Bethe-Salpeter 
approach. Furthermore, the mass of the pion we reach in our calculations 
is considerably closer to the physical value than in previous work.

The computational effort involved in taking the improvement into account 
is small. Since it guarantees elimination of ${\mathcal{O}}(a)$ 
discretisation errors to all orders in the coupling constant, use of 
this method in future work seems logical in pushing the calculations 
further towards the physical limit.

\section*{Acknowledgements}
The authors thank R. Woloshyn for stimulating discussions. The work of
J.v.d.H., J.H.K. and M.L. is part of the research program of the
Foundation for Fundamental Research of Matter (FOM) and the National
Organisation for Scientific Research (NWO).  The research of E.L. is 
partly supported by Deutsche Forschungsgemeinschaft (DFG)  under grant 
FOR 339/2-1. The computations were performed at the Konrad-Zuse Zentrum, 
Berlin, and the John von Neumann-Institut f\"ur Computing (NIC),
J\"ulich.

\end{document}